\documentclass[12pt,russian]{article}
\usepackage[active]{srcltx}
\usepackage{epsfig}
\usepackage{graphicx}
\usepackage{cite}
\usepackage[russian]{babel}
\usepackage[cp866]{inputenc}
\newcommand{\av}{\mbox{\boldmath $a$}}

\newcommand{\mb}{\mbox{\boldmath $\mu$}}

\newcommand{\sbb}{\mbox{\boldmath $s$}}

\newcommand{\bb}{\mbox{\boldmath $b$}}

\newcommand{\B}{\mbox{\boldmath $B$}}

\newcommand\fr{\displaystyle\frac}

\newcommand{\htts}{\mbox{\boldmath$\hat{t}\kern1pt$}}

\newcommand{\qm}{quantum mechanics}
\newcommand\lt{\left}
\newcommand\rt{\right}

\newcommand{\Sh}{Schr\"odinger equation}

\newcommand{\eref}[1]{Eq. (\ref{#1})}
\begin{document}
\begin{center}
{\bf Contradiction of the DENSITY MATRIX notion in \qm }

V.K.Ignatovich

FLNP, JINR, Dubna, Russia.

Abstract \end{center}

It is shown that description of a nonpolarized neutron beam by
density matrix is contradictory. Density matrix is invariant with
respect to choice of quantization axis, while experimental devices
can discriminate between different quantization axes.

\bigskip

\hfill\parbox{9.3cm}{\it ---The statement in the title is
evidently wrong at the modern level of knowledge. The density
matrix is not an auxiliary construction but a result of basic
concepts of quantum mechanics. More over experience with this
notion is huge and convincing. As for the given paper, no doubt it
contains an error. It is in inexactitude of wordings and
reasonings. The problem is only how to find this error. However it
is a task for the author.}

\hfill{ Referee of JETP Lett.}

\section{Introduction}

The main notions in nonrelativistic quantum mechanics are the \Sh\
and wave function $|\psi\rangle$. The density matrix is an
artificial construction, which, as will be shown bellow, can be
contradictory. We will consider the simplest case of the density
matrix, describing a monochromatic nonpolarized neutron beam.

A monochromatic non polarized neutron beam is characterized by the
density matrix
\begin{equation}\label{dm}
\rho=\fr1{2}\Big(|u\rangle\langle u|+|d\rangle\langle d|\Big),
\end{equation}
which is one half of the unit matrix. The states $|u,d\rangle$
correspond to wave functions for neutrons polarized along and
opposite some direction, which is known as quantization axis. The
choice of the quantization axis, however, is not important,
because the density matrix \eref{dm} is invariant with respect to
such a choice. Indeed, if one chooses the quantization axis along
some unit vector $\av$, then the matrix \eref{dm} becomes
\begin{equation}\label{2}
\rho=\fr1{2}\Big(|\av\rangle\langle \av|+|-\av\rangle\langle
-\av|\Big).
\end{equation}
If one chooses another axis $\bb$, then, since
\begin{equation}\label{3}
|\av\rangle=\alpha|\bb\rangle+\beta|-\bb\rangle,\qquad
|-\av\rangle=\alpha^*|-\bb\rangle-\beta^*|\bb\rangle,
\end{equation}
where $|\alpha|^2+|\beta|^2=1$, one obtains
\begin{equation}\label{4}
\rho=\fr1{2}\Big(\lt[\alpha|\bb\rangle+\beta|-\bb\rangle\rt]\lt[\alpha^*\langle\bb|+\beta^*\langle-\bb|\rt]
+$$
$$+\lt[\beta^*|\bb\rangle-\alpha^*|-\bb\rangle\rt]\lt[\beta\langle\bb|-\alpha\langle-\bb|\rt]\Big)=$$
$$=
\fr1{2}\Big(|\bb\rangle\langle \bb|+|-\bb\rangle\langle
-\bb|\Big).
\end{equation}
For instance, if $\av$ is along $y$ axis, and $\bb$ is along
$z$-axis, one has
\begin{equation}\label{5}
|y\rangle=\fr1{\sqrt2}{1\choose
i}=\fr1{\sqrt2}\lt(|z\rangle+i|-z\rangle\rt),\qquad
|-y\rangle=\fr1{\sqrt2}{i\choose1}=\fr1{\sqrt2}\lt[|-z\rangle+i|z\rangle\rt],
\end{equation}
and
\begin{equation}\label{6}
\rho=\fr1{2}\Big(|+z\rangle\langle+z|+|-z\rangle\langle-z|\Big)=\fr1{2}\Big(|+y\rangle\langle+y|+|-y\rangle\langle-y|\Big).
\end{equation}
So two axes are equivalent for the density matrix.  However these
axes can be discriminated by an experimental equipment, and our
goal is to show how it is possible. To achieve it let's first show
how one can find polarization direction of a polarized beam.

\section{A method for polarization direction measurement}

The principle is based on an effect known in neutron
optics~~\cite{vf,ga,uig}, and is related to spin flip with the
help of a resonant radio frequency (rf) spin-flipper. Such a
spin-flipper is a coil with a permanent magnetic field $\B_0$ and
perpendicular to it rotating counterclockwise rf-field
\begin{equation}\label{rf}
\B_{rf}=b\Big(\cos(\omega t),\sin(\omega t).0\Big),
\end{equation}
where $\omega=2\mu B_0/\hbar$, and $\mu$ is magnetic moment of the
neutron, which is aligned oppositely to the neutron spin $\sbb$.
Direction of $\B_0$ can be accepted as the quantization z-axis.
Interaction of neutrons with such a flipper can be solved exactly
and analytically, and the solution can be explained as
follows~\cite{uig}.

The neutron interaction with magnetic field is described by the
potential $-\mb\cdot\B_0$. Therefore neutrons in the state
$|z\rangle$ entering the field $\B_0$ are decelerated because the
field in this case creates a potential barrier of height $\mu
B_0$.

Inside the flipper the rf-field turns the spin down, i.e.
transforms the state $|z\rangle$ into $|-z\rangle$. In this state
the interaction $-\mb\cdot\B_0$ becomes negative, so the potential
barrier transforms into potential well of depth $\mu B_0$.
Therefore after exit from the flipper and its magnetic field
$\B_0$ the neutron decelerates once again. In total the neutron
energy after transmission through the spin flipper decreases by
amount $2\mu B_0$, which means emission of an rf quantum:
$\hbar\omega=2\mu B_0$. The wave functions before and after spin
flipper are
\begin{equation}\label{rf2}
|\psi_{in}(x,t)\rangle=\exp(ikx-i\Omega t)|z\rangle,
\end{equation}
\begin{equation}\label{rf3}
|\psi_{out}(x,t)\rangle=\exp(ik_-(x-D)-i(\Omega-\omega)
t)|-z\rangle,
\end{equation}
respectively. Here $x$ is the axis of propagation, $D$ is
thickness of the spin-flipper, $k$ is initial wave number,
$\Omega=\hbar k^2/2m$, $m$ is the neutron mass, and
$k_-=\sqrt{k^2-2m\omega/\hbar}$. If the incident neutron has the
state $|-z\rangle$ it accelerates, and after spin-flipper has
energy larger than original one by the amount $2\mu B_0$, which
means absorbtion of an rf quantum: $\hbar\omega=2\mu B_0$. The
wave functions before and after spin flipper in this case are
respectively
\begin{equation}\label{rf4}
|\psi_{in}(x,t)\rangle=\exp(ikx-i\Omega t)|-z\rangle,
\end{equation}
\begin{equation}\label{rf5}
|\psi_{out}(x,t)\rangle=\exp(ik_+(x-D)-i(\Omega+\omega)
t)|z\rangle,
\end{equation}
where $k_+=\sqrt{k^2+2m\omega/\hbar}$.

If the incident neutron has a polarization
$|\xi\rangle=\alpha|z\rangle+\beta|-z\rangle$, its wave function
before and after spin flipper are respectively
\begin{equation}\label{rf6}
|\psi_{in}(x,t)\rangle=\exp(ikx-i\Omega t)(\alpha|z\rangle+\beta|-z\rangle),
\end{equation}
\begin{equation}\label{rf7}
|\psi_{out}(x,t)\rangle=\alpha\exp(ik_-(x-D)-i(\Omega-\omega)
t)|-z\rangle+$$
$$+\beta
\exp(ik_+(x-D)-i(\Omega+\omega) t)|z\rangle.
\end{equation}
The spin arrow of this
state represents a rotating spin wave propagating along $x$-axis.

Let's put at some position $x=x_0$ an analyzer, which transmits
only neutrons polarized along $y$-axis. Since
\begin{equation}\label{5}
|+z\rangle=\fr1{\sqrt2}(|+y\rangle-i|-y\rangle),\qquad
|-z\rangle=\fr 1{i\sqrt2}(|+y\rangle+i|-y\rangle),
\end{equation}
where $|\pm y\rangle$ denote states with polarization along and
opposite $y$ axis, the neutron state \eref{rf7} after the analyzer
is
\begin{equation}\label{6}
|\psi_{+y}(x_0,t)\rangle=$$ $$\fr {|+y\rangle}{i\sqrt2}\lt(\alpha
e^{ ik_-(x_0-D)-i(\Omega-\omega)t}+i\beta e^{
ik_+(x_0-D)-i(\Omega+\omega)t}\rt),
\end{equation}
and intensity of the neutron beam after the analyzer at some
position $x_0$ is
\begin{equation}\label{7}
I_{+y}(x_0,t)=\fr12\lt[|\alpha|^2+|\beta|^2+2|\alpha\beta|\cos(\varphi+2\omega
t)\rt],\end{equation}
 where $\varphi$ is some phase. We see that
 the beam has density modulation with time, and visibility of the modulation
 \begin{equation}\label{8}
 V=\fr{2|\alpha\beta|}{|\alpha|^2+|\beta|^2}=\fr{2|\alpha/\beta|}{1+|\alpha|^2/|\beta|^2}
\end{equation}
determines ratio $|\alpha/\beta|$ and, therefore, the polar angle
of the incident neutron spin arrow with respect to $z$-axis. If
$\alpha$ or $\beta$ are zero, i.e. incident neutron is polarized
along or opposite spin-flipper axis, oscillations are absent.

\section{An experimental possibility for discrimination between $z$ and $y$ quantization axes}

Now let's suppose that quantization axis is directed along
$y$-axis. It means that the number $N_+$ of particles in the state
$|+y\rangle$ is the same as the number $N_-$ in the state
$|-y\rangle$. Since $|\pm y\rangle=(|\pm z\rangle+ i|\mp
z\rangle)/\sqrt2$, we have according to \eref{6} the intensities
after $y$-analyzer for two incident components $|\pm y\rangle$
measured by a detector at some position $x_0$ to be
\begin{equation}\label{7a}
I_{+y}^\pm(x_0,t)=\fr{N_\pm}2\lt[1\pm\cos(2\omega
t)\rt],\end{equation} where upper index points out what was the
incident component, and for simplicity we put the phase $\varphi$
in \eref{7a} to zero, because it is the same for all the
particles.

The sum of averaged over time two intensities is a constant
\begin{equation}\label{7a1}
\langle I_{+y}(t)\rangle=\langle I^+_{+y}(t)\rangle+\langle
I^-_{+y}(t)\rangle=$$ $$=\fr{\langle
N_+\rangle}2\lt[1+\cos(2\omega t)\rt]+\fr{\langle
N_-\rangle}2\lt[1-\cos(2\omega t)\rt]=N_0,\end{equation} where
$N_0=\langle N_+\rangle=\langle N_-\rangle$.

However besides the average value there are also fluctuations of
neutron count rate. We can naturally suppose that the fluctuations
of two incident spin components are independent, and obey the
Poisson statistics. Then fluctuations of neutron flux density
after $y$-analyzer will be

\begin{equation}\label{7a2}
\langle|\delta I_{+y}(t)|^2\rangle=\langle|\delta
I^+_{+y}(t)|^2\rangle+\langle|\delta I^-_{+y}(t)|^2\rangle=$$
$$=\Big\langle\fr{\delta N_+}2\lt[1+\cos(2\omega
t)\rt]\Big\rangle^2+\Big\langle\fr{\delta N_-}2\lt[1-\cos(2\omega
t)\rt]\Big\rangle^2=\fr{N_0}{2}(1+\cos^2(2\omega
t)).\end{equation}

To see these oscillations one should divide the period
$T=\pi/2\omega$ over $N$ small intervals $\Delta T=T/N$ and sum
the value
\begin{equation}\label{7aa2}
\fr{\langle|\delta
I_{+y}(t_n)|^2\rangle}{N_0}=\fr{1}2\lt[1+\cos^2(t_n/
T)\rt],\end{equation} at $t_n=n\Delta T$ over many periods $T$.

This way one can discriminate between two quantization axes $z$,
and $y$. Therefore these quantization axes are not equivalent,
whereas according to density matrix expression they are absolutely
equivalent. This is the contradiction we wanted to point to.

\section{Conclusion}

The main element of \qm\ is a wave function, and corresponding to
it a pure state. If one has an ensemble of particles with
different pure states, and the distribution of different states is
characterized by probabilities, one must calculate a process with
pure states and then average over probabilities. This is the way
neutron scattering cross sections are calculated. First they are
calculated for a pure state of an incident plain wave, and then
the obtained cross section is averaged over probability
distribution of the incident plain waves. Of course the density
matrix also can be useful, but because of discovered
contradiction, one must be very careful with it.

\end{document}